\documentstyle[preprint,aps,eqsecnum]{revtex}
\newcommand{\lrvec}[1]{\stackrel{\leftrightarrow}{#1} }
\begin{document}
\draft
\bigskip
\title{ Color-octet mechanism in \\ $\gamma + p \rightarrow J/\psi + X$}
\author{ Pyungwon Ko$^{1,}$\footnote{ pko@phyb.snu.ac.kr}}
\address{$^{1}$Department of Physics, Hong-Ik University,
Seoul 121-791, Korea }
\author{Jungil Lee$^{2,}$\footnote{jungil@phyb.snu.ac.kr}
    and H.S. Song$^{2,}$\footnote{hssong@physs.snu.ac.kr} }
\address{
$^{2}$Center for Theoretical Physics and Department of Physics 
\\ Seoul National University,
Seoul 151-742, Korea  }
\tighten
\maketitle
\begin{abstract}
Photoproduction of $J/\psi$ is considered including the color-octet 
contributions from the various partial wave states,  
$^{2S+1}L_J$ = $^1S_0$, $^3S_1$ and $^3P_J$.
The production cross section depends on three new nonperturbative parameters
defined in NRQCD, called the color-octet matrix elements. 
Using the color-octet  matrix elements determined by fitting the $J/\psi$
production at the Tevatron, we find that the color-octet 
($c\bar{c})_{8} ({^1S_0}~{\rm and ~} ~{^3P_J})$ contributions to the $J/\psi$
photoproductions at the fixed target experiments  and HERA  
are too large compared to the data on $\sigma (\gamma + p \rightarrow J/\psi 
+X)$ in the forward direction, the $z$ distribution of $J/\psi$. 
The $P_T^2$ distribution of $J/\psi$ 
and the total inelastic  $J/\psi$ production rate as a function of 
$\sqrt{s_{\gamma p}}$ are predicted including color-octet contributions.  
We also briefly digress on the $B \rightarrow  J/\psi + X$ and observe the 
similar situation.  This may be an indication that
the color-octet matrix elements determined from the $J/\psi$ production 
at the Tevatron, especially $\langle 0 | {\cal O}_{8}^{\psi} ({^1S_0}) | 0 
\rangle$ and $\langle 0 | {\cal O}_{8}^{\psi} ({^3P_J}) | 0 \rangle$,
might have  been  overestimated by an order of magnitude.  
\end{abstract}
\narrowtext
 \tighten
 \section{Introduction}
 \label{sec:intro}
In the conventional approach, the inelastic (inclusive) $J/\psi$ 
photoproduction  
has been studied in the framework of perturbative QCD (PQCD) and the 
color singlet model \cite{berger}.  
In this model, one considers $\gamma + g \rightarrow 
J/\psi + g$ which can produce high $p_T$ $J/\psi$'s at the $e p $ or 
$\gamma p $ collision.   However, the same approach, when applied to the 
$J/\psi$ or $\psi^{'}$ production at the Tevatron, severely underestimates
the productions rate \cite{mangano}.  
In order to reconcile the data and PQCD predictions,
a new mechanism for heavy quarkonium productions has been suggested 
\cite{fleming},  the color-octet gluon fragmentation into $J/\psi$.  
Also, the color-octet mechanism in heavy quarkonium  productions at hadron 
colliders through the color-octet $(c \bar{c})_8$ pair in various partial 
wave  states ${^{2S+1}L_J}$  
has been considered beyond the color-octet gluon fragmentation approach 
\cite{cho}, \cite{cho2}. 
The main motivation is that inclusive $\Upsilon$ productions at the 
Tevatron also show the excess of the data over theoretical estimates of 
the productions based on PQCD and the color singlet model \cite{vaia}.
Here, the $p_T$ of the $\Upsilon$ is not that high so that the gluon
fragmentation picture may not be a good approximation any more.    
In Refs.~\cite{cho} and  \cite{cho2}, 
a large class of color-octet diagrams has been 
considered which can contribute to the $J/\psi$ production at hadron 
colliders.  At the partonic level, there appear new $2 \rightarrow 1$ 
subprocesses :
\begin{eqnarray}
q \bar{q} & \rightarrow & (c\bar{c})(^{3}S^{(8)}_{1}), 
\\
g g & \rightarrow & (c\bar{c})({^{1}S^{(8)}_{0}}~{\rm or~} {^{3}P^{(8)}_{J}}),
\end{eqnarray}
at the short distance scale, and the subsequent evolution of the 
$(c\bar{c})_{8}({^{2S+1}L_{J}})$ object into a physical $J/\psi$ by 
absorbing/emitting soft gluons at the long distance scale.   
The short distance process can be calculated using PQCD in powers of 
$\alpha_s$, whereas  the long distance part is treated as a new parameter
$\langle 0 | O_{8}^{\psi} ({^{2S+1}L_{J}}) | 0 \rangle$ 
which characterizes the probability that the color-octet $(c\bar{c}) 
({^{2S+1}L_{J}})$ state evolves into a physical $J/\psi$.  
By fitting the $J/\psi$ production at the Tevatron using the usual 
color-singlet production and the cascades from $\chi_{c}(1P)$ and the 
color-octet contribution,  the authors of Ref.~\cite{cho2} determined 
\begin{eqnarray}
\langle 0 | O_{8}^{\psi} (^{3}S_{1}) | 0 \rangle   &=&   (6.6 \pm 2.1)
\times 10^{-3}~{\rm GeV}^3,
\\
\frac{\langle 0|{\cal O}_{8}^{\psi}({^3P_{0}})|0\rangle}{M_c^2}
     +\frac{\langle 0|{\cal O}_{8}^{\psi}({^1S_{0}})|0\rangle}{3}
&=&(2.2\pm 0.5)\times 10^{-2}~{\rm GeV}^3
\end{eqnarray}
for $M_{c} = 1.48$ GeV.
Although the numerical values of two matrix elements $\langle 0|
{\cal O}_{8}^{\psi}(^3P_0)|0\rangle$ and $\langle 0|{\cal O}_{8}^{\psi}
(^1S_0)|0\rangle$ are not separately known in Eq.~(1.4), one can still
extract some useful information from it. Since both of the color octet 
matrix elements in Eq.~(1.4) are positive definite, one has
\begin{eqnarray}
0 < \langle 0|{\cal O}_{8}^{\psi}({^1S_0})|0\rangle < (6.6 \pm 1.5) 
\times 10^{-2}~{\rm GeV}^3,
\\
0 < { \langle 0|{\cal O}_{8}^{\psi}({^3P_0})|0\rangle \over M_{c}^2} 
<  (2.2 \pm 0.5) \times 10^{-2}~{\rm GeV}^3.
\end{eqnarray}
These inequalities can provide us with some predictions on various 
quantities related with inclusive $J/\psi$ productions in other 
high energy processes, which enables us to test the idea of color-octet
mechanism.

Since the color-octet mechanism in heavy quarkonium production is  a 
new idea proposed in order to resolve the $\psi^{'}$ anomaly at the 
Tevatron, it is 
important to test this idea in other high energy processes with 
inclusive heavy quarkonium productions. Up to now, the following 
processes have been considered : $J/\psi$ production at the Tevatron
and fixed target experiments \cite{cho} \cite{cho2} \cite{fleming1}, 
spin alignment of the color-octet produced 
$J/\psi$ \cite{wise}, the polar angle distribution of the $J/\psi$ in the 
$e^+ e^-$ annihilations into $J/\psi +X$ \cite{chen},
inclusive $J/\psi$ production in 
$B$ meson decays \cite{ko}  and the $Z^0$ decays at LEP \cite{cheung}.   
These processes
also  depend on the aforementioned  three color-octet matrix elements 
in different
combinations from (1.4).   Thus, one can check if the color-octet mechanism
provides reasonable agreements between PQCD and the experimental data
on inclusive $J/\psi$ production rates from these processes.

In the above list of various inclusive $J/\psi$ productions at high energies, 
the $J/\psi$ photoproduction is missing, however.  
It is the purpose of this work to study the color-octet mechanism in 
the $J/\psi$ photoproduction \footnote{While we were finishing this work, we 
received two preprints which discussed the same topic \cite{kramer2} 
\cite{fleming2}. We find our results agree with theirs in the case the direct
comparison is possible.}.

In Sec.~\ref{sec:two}, we demonstrate how to get the inclusive production 
rate of a heavy 
quarkonium in the NRQCD formalism of Bodwin, Braaten and Lepage \cite{bbl}.  
Then, we review briefly the $\gamma g$ fusion into $J/\psi + g$ in the 
color-singlet model in Sec.~\ref{subsec:3a}. Then in Sec.~\ref{subsec:3b}, 
we consider the color-octet subprocesses 
\begin{equation}
\gamma + g \rightarrow (c \bar{c})({^{1}S^{(8)}_{0}} ~{\rm or}~ 
{^{3}P^{(8)}_{J=0,2}}),
\end{equation}
which have  not been included in previous studies.   
The size of these color-octet contributions to the $J/\psi$
photoproductions are  suppressed by $v^4$ 
relative to the color-singlet contributions, but of lower order in 
$\alpha_s$.  This subprocess contributes to the $J/\psi$ photoproduction
in the forward scattering (the elastic peak) with $z \approx 1$ and 
$P_{T}^{2} \approx 0$.  
These color-octet $2 \rightarrow 1$ subprocesses can also contribute to
the $2 \rightarrow 2$ subprocesses through 
\begin{eqnarray}
\gamma+g&\rightarrow&(c\bar{c})(^{1}S^{(8)}_{0}~{\rm or}~^{3}P^{(8)}_{J}) + g,
\\
\gamma+q&\rightarrow&(c\bar{c})(^{1}S^{(8)}_{0}~{\rm or}~^{3}P^{(8)}_{J}) + q.
\end{eqnarray}
These are also resolved photon processes at lower order 
[$O(\alpha \alpha_{s}^2)]$ than the color-singlet model  [$O(\alpha 
\alpha_{s}^3)$]  in the perturbation expansion in $\alpha_s$ :
although the color-octet contributions are suppressed by $v^4$ compared to 
the color-singlet resolved photon process.  Thus, the color-octet $^{1}S_{0}$
and $^{3}P_{J}$ states can contribute to the elastic peak of the 
$J/\psi-$photoproduction as well as contribute to the resolved photon process.
It is quite important to estimate the latter and compare with the resolved 
photon process in the color-singlet model, 
since it is a common statement that $J/\psi-$photoproduction is
a good place to measure the gluon distribution function in a proton.  
We find that the quark contributions are small compared to the gluon
contribution even if we include (1.9).  
When one considers Eqs.~(1.8) and (1.9), one has to consider 
\begin{equation}
\gamma + g \rightarrow (c \bar{c})({^{2S+1}L^{(8)}_{J}}) + g,
\end{equation}
although it is expected to be suppressed relative to the usual $\gamma g$ 
fusion color-singlet diagram by $v^4$. It is gauge invariant by itself, and 
thus can be safely neglected if we wish.  We keep it however, 
in order to be consistent 
in the $\alpha_s$ expansion, and make it sure the $v^2$ scaling rule works 
in this case. All of these color-octet $2 \rightarrow 2$ subprocesses are 
discussed in Sec.~\ref{subsec:3c}. 
Numerical analyses relevant to the fixed target experiments and HERA 
are performed in Sec. ~\ref{subsec:photo}.  
We show that the relations (1.5) and (1.6)
yields too large a cross section for the $J/\psi$ 
photoproduction in the forward direction. They also leads to too rapidly  
growing $d \sigma / dz$ distribution for high $z$ region compared 
to the experimental observations.  
In Sec.~\ref{subsec:btocc}, we briefly digress on 
the $B \rightarrow J/\psi + X$ using the 
factorization formula derived in Ref.~\cite{ko}, and find again that the 
relations (1.5) and (1.6) overestimate the branching ratio for $B \rightarrow 
J/\psi + X$.  All of these seem to indicate that the relations (1.3) and 
(1.4), especially the latter, are probably overestimated by an order of 
magnitude.  This is not surprising at all, since the analyses in 
Ref. ~\cite{cho2} employed the leading order calculations for the color-singlet
parton subprocess for the $J/\psi$ hadroproduction.
We summarize our results and speculate the origins of these overestimates of 
$J/\psi$ photoproductions and $B$ meson decays in Sec.~\ref{sec:five}.  

\vspace{.3in}

\section{NRQCD formalism for heavy quarkonium productions}
\label{sec:two}

To begin, we consider  general methods to get the NRQCD cross section
of the process
$a+b\rightarrow Q\bar{Q}(^{2S+1}L^{(1,8a)}_J)\rightarrow  H$
,where $H$ is the final quarkonium state and $Q\bar{Q}(^{2S+1}L^{(1,8a)}_J)$ 
is the intermediate  $Q\bar{Q}$ pair which  fragments into a specific 
heavy quarkonium state in the long distance scale.
If the on-shell scattering amplitude of the process
${\cal A}(a+b\rightarrow Q+\bar{Q})$ is given, 
we can expand the amplitude in terms of relative momentum $q$ of
the quarks inside the bound state
because the quarks which make the bound state are heavy.
Scattering amplitude of the process  
$a+b\rightarrow Q\bar{Q}(^{2S+1}L^{(1,8a)}_J)(P)$
is given by
\begin{eqnarray}
{\cal A}\left(a+b\rightarrow Q\bar{Q}(^{2S+1}L^{(1,8a)}_J)(P)\right)
&=&
\sum_{L_ZS_Z} \sum_{s_1s_2} \sum_{ij}
\int\frac{d^3\vec{q}}{(2\pi)^32q^0}\delta(q^0-\frac{|\vec{q}|^2}{2M_Q})
Y^*_{LL_Z}(\hat{q})\nonumber\\
&&\times
\langle s_1;s_2|SS_Z\rangle
\langle LL_Z;SS_Z|JJ_Z\rangle
\langle 3i\bar{3}j|1,8a\rangle\nonumber\\
&&\times
{\cal A}(a+b\rightarrow Q^i(\frac{P}{2}+q)+\bar{Q}^j(\frac{P}{2}-q)),
\end{eqnarray}
where the superscript $(1,8a)$ represents the color-singlet or the 
color-octet configuration of the $Q \bar{Q}$ state.  
After  integrating over the relative momentum $q$,
we get
\begin{equation}
{\cal A}\left(a+b\rightarrow Q\bar{Q}(^{2S+1}L^{(1,8a)}_J)\right)
=\sqrt{C_L}{\cal M}^\prime_L
\left(a+b\rightarrow Q\bar{Q}(^{2S+1}L^{(1,8a)}_J)\right),
\end{equation}
where  
\begin{eqnarray}
{\cal M}^\prime_L
\left(a+b\rightarrow Q\bar{Q}(^{2S+1}L^{(1,8a)}_J)\right)
&=&\sum_{L_ZS_Z} \sum_{s_1s_2} \sum_{ij} 
\langle s_1;s_2|SS_Z\rangle
\langle LL_Z;SS_Z|JJ_Z\rangle
\langle 3i\bar{3}j|1,8a\rangle
\nonumber\\
&&\hskip -.5cm
\times
\left\{
\begin{array}{ll}
{\cal A}(a+b\rightarrow Q^i+\bar{Q}^j)|_{q=0}
&(L=S) ,\\
\epsilon^*_{\alpha}(L_Z)
{\cal A}^\alpha(a+b\rightarrow Q^i+\bar{Q}^j)|_{q=0}
&(L=P) ,\\
\epsilon^*_{\alpha\beta}(L_Z)
{\cal A}^{\alpha\beta}(a+b\rightarrow Q^i+\bar{Q}^j)|_{q=0}
&(L=D),
\end{array}
\right.
,
\\
C_L=\frac{M_Q^3}{2q^0(2\pi)^5}\times|\vec{q}|^{2L}
\hskip .3cm
&{\rm and}&
\hskip .3cm
C^{rel}_L=\frac{M_Q^3}{2q^0(2\pi)^5}\times|\vec{q}|^{2L+2}.
\end{eqnarray}
Here, $C^{rel}_L$ is the factor
for the relativistic correction of the $L$ state.
>From now on, amplitudes with Lorentz indices 
mean
\begin{equation}
A^\alpha(P,q)=\frac{\partial}{\partial q^\alpha}A(P,q)
\hskip .5cm
{\rm and}
\hskip .5cm
A^{\alpha\beta}(P,q)=\frac{\partial^2} {\partial q^\alpha\partial q^\beta}
A(P,q).
\end{equation}
The color SU(3) coefficients are given by
\begin{equation}
\langle 3i;\bar{3}j|1\rangle=\delta_{ij}/\sqrt{N_c}
~~{\rm and}~~ 
\langle 3i;\bar{3}j|8a\rangle=\sqrt{2}T^a_{ij}.
\end{equation}
For the case of color singlet state, we can relate the coefficients $C_L$
to the radial wavefunction of the bound state as
\begin{eqnarray}
C_S=\frac{1}{4\pi}|R_S(0)|^2,
~~~
C_P=\frac{3}{4\pi}|R_P^\prime(0)|^2
~~{\rm and}~~
C_D=\frac{15}{8\pi}|R_D^{\prime\prime}(0)|^2.
\end{eqnarray}
Some identities of color matrix trace are useful
in squaring the matrix elements.
\begin{equation}
\begin{array}{rclcrcl}
{\rm Tr}(1)                               &=&+N_c,&~~~&
{\rm Tr}(T^a T^b T^c T^c T^b T^a)         &=&+\frac{(N_c^2-1)^3}{8N_c^2},\\

{\rm Tr}(T^a T^a)                         &=&+\frac{N_c^2-1}{2},&~~~&
{\rm Tr}(T^a T^b T^c T^b T^c T^a)         &=&-\frac{(N_c^2-1)^2}{8N_c^2},\\

{\rm Tr}(T^a T^b T^b T^a)                 &=&+\frac{N_c^2-1}{4N_c},&~~~&
{\rm Tr}(T^a T^b T^c T^a T^b T^c)         &=&+\frac{(N_c^4-1)  }{8N_c^2},\\

{\rm Tr}(T^a T^b T^a T^b)                 &=&-\frac{N_c^2-1}{4N_c},&~~~&
{\rm Tr}(T^a T^b T^c){\rm Tr}(T^a T^b T^c)&=&-\frac{(N_c^2-1)  }{4N_c},\\

{\rm Tr}(T^a T^b){\rm Tr}(T^a T^b)        &=&+\frac{N_c^2-1}{4},&~~~&
{\rm Tr}(T^a T^b T^c){\rm Tr}(T^a T^c T^b)&=&-\frac{(N_c^2-1)(N_c^2-2)}{4N_c}.
\end{array}
\end{equation}
At this stage we can derive the explicit form of the matrix element
${\cal M}^\prime_L $.
In general, the on-shell amplitude can be expressed as
\begin{equation}
\langle 3i\bar{3}j|1,8a\rangle
{\cal A}(a+b\rightarrow Q^i(\frac{P}{2}+q)+\bar{Q}^j(\frac{P}{2}-q))
=\bar{u}(\frac{P}{2}+q;s_1)
{\cal O}(P,q)
v(\frac{P}{2}-q;s_2),
\end{equation}
where ${\cal O}$ is the matrix relevant to the on shell amplitude.
If we introduce the spin projection operator $P_{SS_z}(P,q)$ as
\begin{equation}
P_{SS_z}(P,q)_{ij}\equiv 
\sum_{s_1 s_2}
\langle s_1;s_2|SS_Z\rangle
v_i(\frac{P}{2}-q;s_2) \bar{u}_j(\frac{P}{2}+q;s_1),
\end{equation}
we can simplify the form of the matrix element ${\cal M}^\prime_L$  as
\begin{eqnarray}
{\cal M}^\prime_S&=& {\rm Tr}\left[{\cal O}(P,0) P_{SS_z}(P,0)\right],
\\
{\cal M}^\prime_P&=& \sum_{L_zS_z} \epsilon^{*}_\alpha(L_Z)
                   \langle LL_z;SS_z|JJ_z\rangle
                   {\rm Tr}
                   \left[ {\cal O}^\alpha P_{SS_z}
                         +{\cal O}P^\alpha_{SS_z}
                   \right]_{q=0},
\\
{\cal M}^\prime_D&=& \sum_{L_zS_z} \epsilon^{*}_{\alpha\beta}(L_Z)
                   \langle LL_z;SS_z|JJ_z\rangle
                   {\rm Tr}
                   \left[ {\cal O}^{\alpha\beta} P_{SS_z}
                         +{\cal O}^\alpha P^\beta_{SS_z}
                         +{\cal O}^\beta P^\alpha_{SS_z}
                         +{\cal O} P^{\alpha\beta}_{SS_z}
                   \right]_{q=0}.
\end{eqnarray}
Note that ${\cal O}$ includes the color coefficient 
$\langle 3i\bar{3}j|1,8a\rangle$
and $P_{SS_z}$ possesses the spin coefficient
$\langle s_1;s_2|SS_Z\rangle$.
Expanding ~$P_{SS_z}(P,q)$ to the second  order of the relative momentum
$q$, we get
\begin{eqnarray}
P_{00}(P,0) &=& \frac{1}{2\sqrt{2}}\gamma_5(\not{P}+2M_Q),
\\
P^\alpha_{00}(P,0) 
    &=& \frac{1}{2\sqrt{2}M_Q}\gamma^\alpha\gamma_5\not{P},
\\
P^{\alpha\beta}_{00}(P,0) 
    &=&-\frac{1}{2\sqrt{2}M_Q}g^{\alpha\beta} 
        \gamma_5,
\\
P_{1S_z}(P,0) &=&
\frac{1}{2\sqrt{2}} \not{\epsilon}^*(S_z)(\not{P}+2M_Q),
\\
P^\alpha_{1S_z}(P,0) &=&
\frac{1}{4\sqrt{2}M_Q}
\left[
 \not{\epsilon}^*(S_z)(\not{P}+2M_Q)\gamma^\alpha
+\gamma^\alpha\not{\epsilon}^*(S_z)(\not{P}+2M_Q)
\right],
\\
P^{\alpha\beta}_{1S_z}(P,0)
    &=&-\frac{1}{2\sqrt{2}M_Q}
\left[ g^{\alpha\beta}\not{\epsilon}^*(S_z)
      -\frac{1}{4M_Q}(\not{P}+2M_Q)
      ( {\epsilon}^{*\alpha}(S_z)\gamma^\beta
       +{\epsilon}^{*\beta}(S_z)\gamma^\alpha
      )
\right].
\end{eqnarray}
When $L=P$, we need further relations to get the correct polarization
state of the intermediate state,
\begin{eqnarray}
\sum_{L_ZS_Z}
\epsilon^{*\alpha(L_z)}
\epsilon^{*\beta}(S_z)
\langle 1L_z;1S_z|J=0~J_z=0\rangle&=&\frac{1}{\sqrt{3}}
(-g^{\alpha\beta}+\frac{P^\alpha P^\beta}{M^2}),  \nonumber\\
\sum_{L_ZS_Z}
\epsilon^{*\alpha(L_z)}
\epsilon^{*\beta}(S_z)
\langle 1L_z;1S_z|J=1~J_z\rangle&=&-\frac{i}{\sqrt{2}M}
\epsilon^{\alpha\beta\lambda\kappa}
P_\kappa \epsilon^*_\lambda(J_z),   \\
\sum_{L_ZS_Z}
\epsilon^{*\alpha(L_z)}
\epsilon^{*\beta}(S_z)
\langle 1L_z;1S_z|J=2~J_z\rangle&=&
\epsilon^{*\alpha\beta}(J_z),
\nonumber
\end{eqnarray}
where the polarization vector and symmetric polarization tensor have the 
properties 
\begin{equation}
P_\alpha \epsilon^\alpha(J_z)=0,\hskip .5cm
P_\alpha \epsilon^{\alpha\beta}(J_z)=0,\hskip .5cm
\epsilon^\alpha_\alpha(J_z)=0,\hskip .5cm
\epsilon^{\alpha\beta}(J_z)=\epsilon^{\beta\alpha}(J_z).
\end{equation}

Once the cross section of the on-shell parton level process is calculated,
one can expand it in factorized forms following BBL \cite{bbl} as
\begin{eqnarray}
\hat{\sigma}\left(a+b\rightarrow (Q\bar{Q})_n\right)
&=&
\frac{F_n}{M_Q^{d_n-5}}
\times
\frac{\langle 0|{\cal O}^{Q\bar{Q}}_n|0\rangle}{2J+1},
\\
\hat{\sigma}\left(a+b\rightarrow (Q\bar{Q})_n\rightarrow H+X\right)
&=&
\frac{F_n}{M_Q^{d_n-4}}
\times
\frac{\langle 0|{\cal O}^{H}_n|0\rangle}{2J+1}.
\end{eqnarray}
We use $\hat{\sigma}$ instead of $\sigma$, as a subprocess cross section,
since we will consider $\gamma p$ collision where
the particle $b$ is treated as a parton inside a proton.
The index $n$ denotes the intermediate $Q\bar{Q}$ state 
$^{2S+1}L^{(1,8)}_J$, which may differ from that of $H$.
The factor multiplied to the $H$ production cross section
differs from that of $Q\bar{Q}$ state by unity in mass dimension.
This  makes the long range factor coincide with the conventionally
normalized wave function of the bound state for the color singlet case.
We extracted the factor $1/(2J+1)$ in advance 
to avoid the unnecessary factor
after imposing the heavy quark spin symmetry property as
\begin{eqnarray}
\frac{\langle 0|{\cal O}^{\psi}(^3S^{(1,8)}_1)|0\rangle}{3}
&\rightarrow& {\langle 0|{\cal O}^{\psi}(^1S^{(1,8)}_0)|0\rangle},
\\
\frac{\langle 0|{\cal O}^{\psi}(^3P^{(1,8)}_J)|0\rangle}{2J+1}
&\rightarrow& {\langle 0|{\cal O}^{\psi}(^3P^{(1,8)}_0)|0\rangle}.
\end{eqnarray}
The 4-fermion operator ${\cal O}_n$ with the dimension$d_n$ are defined 
as
\begin{equation}
\begin{array}{ll}
  d_n=6& d_n=8\\
{\cal O}_1(^1S_0)=\psi^{\dag}    \chi\chi^{\dag}    \psi,&
{\cal O}_1(^1P_1)=\psi^{\dag} (-\frac{i}{2}\lrvec{D})\chi
                    \cdot\chi^{\dag} (-\frac{i}{2}\lrvec{D})  \psi,\\
{\cal O}_8(^1S_0)=\psi^{\dag} T^a\chi\chi^{\dag} T^a\psi,&
{\cal O}_1(^3P_0)=
\frac{1}{3}
\psi^{\dag} (-\frac{i}{2}\lrvec{D}\cdot\vec{\sigma})\chi
\chi^{\dag} (-\frac{i}{2}\lrvec{D}\cdot\vec{\sigma})\psi,\\
{\cal O}_1(^3S_1)=\psi^{\dag} \vec{\sigma}   \chi
\cdot\chi^{\dag}\vec{\sigma}    \psi,&
{\cal O}_1(^3P_1)=
\frac{1}{2}
\psi^{\dag} (-\frac{i}{2}\lrvec{D}\times\vec{\sigma})\chi
\cdot
\chi^{\dag} (-\frac{i}{2}\lrvec{D}\times\vec{\sigma})\psi,\\
{\cal O}_8(^3S_1)=\psi^{\dag} \vec{\sigma}T^a\chi
\cdot\chi^{\dag}\vec{\sigma} T^a\psi,&
{\cal O}_1(^3P_2)=
\psi^{\dag} (-\frac{i}{2}\lrvec{D}{^{(i}}\sigma^{j)})\chi
\cdot
\chi^{\dag} (-\frac{i}{2}\lrvec{D}{^{(i}}\sigma^{j)})\psi.
  \end{array}
\end{equation}
and dimension-8 operators related to the relativistic correction are 
\begin{equation}
  \begin{array}{rcl}
{\cal P}_1(^1S_0)&=&
\frac{1}{2}\left[
\psi^{\dag}\chi\chi^{\dag} (-\frac{i}{2}\lrvec{D})^2\psi+{\rm h.c.}
           \right],\\
{\cal P}_1(^3S_1)&=&
\frac{1}{2}\left[
\psi^{\dag}\vec{\sigma}\chi
\cdot
\chi^{\dag} \vec{\sigma}(-\frac{i}{2}\lrvec{D})^2\psi+{\rm h.c.}
           \right],\\
{\cal P}_1(^3S_1,^3D_1)&=&
\frac{1}{2}\left[
\psi^{\dag}{\sigma}^i\chi
\chi^{\dag}{\sigma}^j(-\frac{i}{2})^2
\lrvec{D}{^{(i}}\lrvec{D}{^{j)}}\psi+{\rm h.c.}
           \right].
  \end{array}
\end{equation}
where 
\begin{eqnarray}
\chi^{\dag} \lrvec{D} \psi&\equiv&
\chi^{\dag} (\vec{D} \psi) -(\vec{D}\chi)^{\dag}\psi,\\
A^{(ij)}&=&\frac{1}{2}(A^{ij}+A^{ji})-\frac{1}{3}{\rm Tr}(A)\delta^{ij},
\end{eqnarray}
and $\vec{D}$ is the covariant derivative.
There are  Pauli spinor fields in the previous equations.
$\psi$ annihilates a heavy quark $Q$
and  $\chi$ creates a heavy antiquark $\bar{Q}$.
Color and spin indices on the fields $\psi$,  $\chi$ have been suppressed.

Vacuum expectation values of
the production operators 
${\cal O}^{Q\bar{Q}}_n$ and
${\cal O}^{H}_n$ are 
\begin{eqnarray}
\langle 0| {\cal O}^H_n|0\rangle
&=&
\langle 0|
\chi^{\dag}{\cal K}_n \psi
\left(\sum_X\sum_{m_J}|H+X\rangle\langle H+X|\right)
\psi^{\dag}{\cal K}^\prime_n
\chi
|0\rangle,
\\
\langle 0| {\cal O}^{Q\bar{Q}}_n|0\rangle
&=&
\langle 0|
\chi^{\dag}{\cal K}_n \psi
\left(\sum_{m_J}|Q\bar{Q}(^{2S+1}L^{(1,8)}_J)\rangle
\langle Q\bar{Q}(^{2S+1}L^{(1,8)}_J)|\right)
\psi^{\dag}{\cal K}^\prime_n\chi
|0\rangle \nonumber\\
&=&
(2J+1)
\langle 0| \chi^{\dag}{\cal K}_n \psi 
|Q\bar{Q}(^{2S^\prime+1}L^{(8)}_0)\rangle
\langle Q\bar{Q}(^{2S^\prime+1}L^{(1,8)}_0)
| \psi^{\dag}{\cal K}^\prime_n \chi |0\rangle
\\
&=&
(2J+1)
\langle 0| {\cal O}^{Q\bar{Q}}
(^{2S^\prime+1}L^{(1,8)}_0)|0\rangle.   \nonumber
\end{eqnarray}
The factors ${\cal K}_n$ and ${\cal K}^\prime_n$ are 
products of a color matrix,  a spin matrix
and a polynomial in the covariant derivative $\lrvec{D}$ and other fields
, which are same with those of ${\cal O}_n$.
According to the heavy quark spin symmetry,
$^{2S+1}L^{(1,8)}_J$ 
state has the same properties with another state
$^{2S^\prime+1}L^{(1,8)}_0$(with the same $L$)
except for the $m_J$ multiplicity factor $2J+1$, which appears
in the last equation.

Let us explain the process to derive the short distance coefficients
$F_n$.
Regardless of the onshell scattering amplitude $a+b\rightarrow Q+\bar{Q}$,
the intermediate bound state ${Q\bar{Q}}(^{2S+1}L^{(1,8)})$ production
cross sections have common factor
$\langle 0|{\cal O}_n^{Q\bar{Q}}|0\rangle$.
We can obtain the matrix elements by using the intermediate
state ket
\begin{eqnarray}
|Q\bar{Q}(^{2S+1}L^{(1,8a)}_J)(P)\rangle
&=&
\sum_{L_ZS_Z} \sum_{s_1s_2} \sum_{ij}
\int\frac{d^3\vec{q}}{(2\pi)^32q^0}\delta(q^0-\frac{|\vec{q}|^2}{2M_Q})
Y^*_{LL_Z}(\hat{q}) \nonumber\\
&\times&
\langle s_1;s_2|SS_Z\rangle
\langle LL_Z;SS_Z|JJ_Z\rangle
\langle 3i\bar{3}j|1,8a\rangle
|Q^i(\frac{P}{2}+q)\bar{Q}^j(\frac{P}{2}-q)\rangle
\end{eqnarray}
as
\begin{equation}
\frac{\langle 0| {\cal O}^{Q\bar{Q}}_n|0\rangle}{2J+1} 
=C_L\times C_n,\\
\end{equation}
where
\begin{equation}
C_n=\left\{
\begin{array}{cl}
2N_c    &({\rm color~~ singlet})\\
N_c^2-1 &({\rm color~~ octet})
\end{array}
\right.
.
\end{equation}
The cross section of the process 
$a(p_1)+b(p_2)\rightarrow (Q\bar{Q})_n(P)$
(with  $n$ representing the partial wave and the color quantum numbers
of $Q \bar{Q}$)  
is given by
\begin{eqnarray}
\hat{\sigma}(a(p_1)+b(p_2)&\rightarrow& (Q\bar{Q})_n(P))\nonumber\\
&=&
\frac{1}{2\hat{s}}\int\frac{d^3\vec{P}}{(2\pi)^32E_P}
(2\pi)^4\delta^{(4)}(P-p_1-p_2)
\overline{\sum}|{\cal A}\left(a+b\rightarrow (Q\bar{Q})_n\right)|^2
\nonumber\\
&=& \hat{\sigma}^\prime_n
\times
C_L\times
C_n
\nonumber\\
&=&\hat{\sigma}^\prime_n
\times
\frac{\langle 0| {\cal O}^{Q\bar{Q}}_n|0\rangle}{2J+1},
\end{eqnarray}
where
\begin{equation}
\hat{\sigma}^\prime_n
=
\frac{1}{C_n}
\frac{\pi}{\hat{s}}\delta(\hat{s}-M_P^2)
\overline{\sum}|{\cal M}_L^\prime(a+b\rightarrow (Q\bar{Q})_n)|^2,
\end{equation}
$n$ represents the partial wave (${^{2S+1}L_J}$) and the color quantum numbers
of $Q \bar{Q}$
and $\hat{s}$ is the invariant mass of the initial particles $a$ and $b$.
Then, the bound state cross section
and the short distance coefficients $F_n$ are given by 
\begin{eqnarray}
\hat{\sigma}(a(p_1)+b(p_2)\rightarrow (Q\bar{Q})_n\rightarrow H+X)
&=&\frac{\hat{\sigma}^\prime_n}{M_Q}
\times
\frac{\langle 0| {\cal O}^{H}_n|0\rangle}{2J+1}
=\frac{\hat{\sigma}^\prime_n M_Q^{d_n-5}}{M_Q^{d_n-4}}
\times
\frac{\langle 0| {\cal O}^{H}_n|0\rangle}{2J+1},
\\
F_n&=&\hat{\sigma}^\prime_n\times M_Q^{d_n-5}.
\end{eqnarray}
In case of $\gamma p$ scattering,
we should convolute the above result with the parton structure functions 
to get the cross section :
\begin{equation}
\sigma(a(p_1)+b(p_2)\rightarrow (Q\bar{Q})_n\rightarrow H+X)
=\frac{1}{M_Q}\sum_{b}\sigma^\prime_n(b)
\times
\frac{\langle 0| {\cal O}^{H}_n|0\rangle}{2J+1},
\end{equation}
where
\begin{eqnarray}
\sigma_n^\prime(b)
&=&\int dx f_{b/p}(x)\hat{\sigma}^\prime_n\nonumber\\
&=&\frac{\pi}{16C_nM_Q^4}\left[xf_{b/p}(x)\right]_{x=4M_Q^2/s}
   \overline{\sum}|{\cal M}_L^\prime(a+b\rightarrow (Q\bar{Q})_n)|^2.
\end{eqnarray}
For the case of $2\rightarrow 2$ process,
we need to modify the formulae only a little.
The quarkonium $H$ photoproduction cross section
via $2\rightarrow 2$ subprocess
$a(p_1)+b(P_2)/p\rightarrow (Q\bar{Q})_n(P)+c(p_3) \rightarrow  H+ X $,
where $b$ is a parton of the initial proton is given by
\begin{eqnarray}
&d\sigma&\left(
a(p_1)+b(P_2)/p\rightarrow (Q\bar{Q})_n(P)+c(p_3)
             \rightarrow  H+ X
\right)
\nonumber\\
&&\hskip 1cm=
\frac{1}{C_nM_Q}\cdot
\frac{1}{16\pi\hat{s}^2}
\overline{\sum}
|{\cal M}^\prime
\left(
(p_1)+b(P_2)\rightarrow (Q\bar{Q})_n(P)+c(p_3)
\right)|^2
\frac{xf_{b/p}(x)}{z(1-z)}dzdP_T^2.
\end{eqnarray}
For example, if we consider the $J/\psi$ production
via 
$^1S_0^{(8)}$, $^3S_1^{(8)}$, $^3P_0^{(8)}$, 
$^3P_1^{(8)}$ and  $^3P_2^{(8)}$
intermediate states, we get  
\begin{eqnarray}
\hat{\sigma}(H)&=&\frac{1}{M_Q}
\left(
\hat{\sigma}^\prime(^1S_0^{(8)})
\times
\langle 0|{\cal O}^{\psi}(^1S_0^{(8)})|0\rangle
+\hat{\sigma}^\prime(^3S_1^{(8)})
\times
\frac{\langle 0|{\cal O}^{\psi}(^3S_1^{(8)})|0\rangle}{3}
\right.
\nonumber\\
&&
\hskip 1cm
\left.
+\sum_{J}
\hat{\sigma}^\prime(^3P_J^{(8)})
\times
\frac{\langle 0|{\cal O}^{\psi}(^3P_J^{(8)})|0\rangle}{2J+1}
\right)
\nonumber\\
&=&
\frac{1}{M_Q}
\left[
\langle 0|{\cal O}^{\psi}(^1S_0^{(8)})|0\rangle
\times
\left(\hat{\sigma}^\prime(^1S_0^{(8)}) 
     +\hat{\sigma}^\prime(^3S_1^{(8)})\right)
     +\langle 0|{\cal O}^{\psi}(^3P_0^{(8)})|0\rangle
\times
\sum_{J}
\hat{\sigma}^\prime(^3P_J^{(8)})
\right].
\label{eq:sigma}
\end{eqnarray}

\vspace{.3in}

\section{$J/\psi$ photoproduction subprocesses}
\label{sec:three}

\subsection{Color-singlet contributions}
\label{subsec:3a}

The inelastic $J/\psi-$photoproduction has long been studied in the
framework of PQCD and the color-singlet model \cite{berger} \cite{jung}.  
The lowest order
subprocess at the parton level for $\gamma + p \rightarrow J/\psi + X$
is the $\gamma-$gluon  fusion at the short distance scale (Fig.~1),
\begin{equation}
\gamma + g \rightarrow (c\bar{c})(^{3}S^{(1)}_{1}) + g,
\end{equation}
followed by the long distance process
\begin{equation}
(c\bar{c})(^{3}S^{(1)}_{1}) \rightarrow J/\psi,
\end{equation}
at the  order of $O(\alpha \alpha_{s}^{2} v^{3})$ in the nonrelativistic 
limit.  
Thus, the production cross section is proportional to the gluon distribution
inside the proton.
This is why  the $J/\psi-$photoproduction has been advocated as a clean
probe for the gluon structure function of a proton in the color-singlet 
model.
Without further details, we show the lowest order color-singlet 
contribution to $J/\psi$
photoproduction through $\gamma-$gluon fusion in the nonrelativistic limit :
\begin{equation}
\overline{| {\cal M}(\gamma g \rightarrow J/\psi g) |^{2}}
 =  {\cal N}_1~{{ \hat{s}^{2} (\hat{s}-4 M_{c}^{2})^{2} + \hat{t}^{2}
( \hat{t} - 4 M_{c}^{2})^{2} + \hat{u} ( \hat{u} - 4 M_{c}^{2} )^{2} }
\over { (\hat{s}-4 M_{c}^{2})^{2} ( \hat{t} - 4 M_{c}^{2})^{2} ( \hat{u} -
4 M_{c}^{2} )^{2}  }},
\end{equation}
where 
\begin{equation}
\begin{array}{ccccl}
    z & = &  \frac{E_\psi}{E_\gamma}|_{\rm lab}&=&
             {p_{N} \cdot P \over p_{N} \cdot k},
\\
\hat{s} &  = & ( k + q_{1} )^{2} &=& x s,
\\
\hat{t} &  = & ( P - k)^{2} &=& (z-1) \hat{s}.
\end{array}
\end{equation}
The overall normalization ${\cal N}_1$ is defined as
\begin{equation}
{\cal N}_1 = {32 \over 9}~( 4 \pi \alpha_{s})^{2} (4 \pi \alpha) e_{c}^{2}
~M_{c}^{3} G_{1}(J/\psi).
\end{equation}
The parameter $G_1 (J/\psi)$, which is defined as 
\begin{equation}
G_{1}( J/\psi) =  {\langle J/\psi | {\cal O}_{1} ({^{3}S_{1}}) | J/\psi 
\rangle  \over M_c^2}
\end{equation}
in the NRQCD, is proportional to the probability that a color-singlet
$c \bar{c}$ pair in the $^{3}S^{(1)}_{1}$ partial wave state to form a 
physical  $J/\psi$ state.  It is related with the leptonic decay via
\begin{equation}
\Gamma ( J/\psi \rightarrow l^{+} l^{-} ) = {2 \over 3}~\pi e_{c}^{2} 
\alpha^{2}~G_{1}(J/\psi),
\end{equation}
to the lowest order in $\alpha_s$.   From
the measured leptonic decay rate of $J/\psi$, one can extract
\begin{equation}
G_{1}(J/\psi) \approx 106~~{\rm MeV},
\end{equation}
Including the radiative corrections of $O(\alpha_{s})$ with $\alpha_{s}
(M_{c}) =0.27$, it is increased to $\approx 184$ MeV.
Relativistic corrections tend to increase $G_{1} (J/\psi)  $ further to
$\sim 195$ MeV \cite{ko}. 

The partonic cross section for $\gamma + a \rightarrow J/\psi + b$
is given by
\begin{equation}
{d\hat{\sigma} \over d\hat{t}} = {1 \over 16 \pi \hat{s}^2}~
\overline{\sum}
| {\cal M}(\gamma + a \rightarrow J/\psi + b) |^{2}.
\end{equation}
The double differential cross section is
\begin{equation}
{d^{2}\sigma \over dz dP_{T}^2} (\gamma + p (p_{N}) \rightarrow
J/\psi (P, \epsilon) + X)
= {x g(x,Q^{2}) \over z (1 -z)}~{1 \over 16 \pi \hat{s}^2}~
\overline{\sum}| {\cal M} |^2
(\hat{s}, \hat{t}),
\end{equation}
where
\begin{equation}
x  =  \frac{\hat{s}}{s} = 
{1\over z s}~\left[ M_{\psi}^{2} + {P_{T}^{2} \over 1-z } \right].
\end{equation}
One has  the following constraints for $x,z,t$ and $P_{T}^2$ :
\begin{eqnarray}
{M_{\psi}^{2} \over s}    < x < 1,
\\
-(\hat{s} - M_{\psi}^{2}) \leq \hat{t} (=t) \leq 0,
\\
M_{\psi}^{2} \leq {M_{\psi}^{2} \over z} + {P_{T}^{2} \over z(1-z)}
\leq s.
\end{eqnarray}
The $z$ and $P_{T}^2$ distributions can be obtained in the following manner :
\begin{eqnarray}
\frac{d\sigma}{dz}&=&
\int^{(1-z)(zs-M_\psi^2)}_0\frac{d^2\sigma}{dzdP_T^2}dP_T^2,
\\
\frac{d\sigma}{dP_T^2}&=&
\int^{z_{\rm max}}_{z_{\rm min}}\frac{d^2\sigma}{dzdP_T^2}dz,
\\
z_{\rm max}&=&\frac{1}{2s}
\left(
s+M_\psi^2+\sqrt{(s-M_\psi^2)^2-4sP_T^2}
\right),
\\
z_{\rm min}&=&\frac{1}{2s}
\left(
s+M_\psi^2-\sqrt{(s-M_\psi^2)^2-4sP_T^2}
\right).
\end{eqnarray}

There are two kinds of corrections to the lowest order result in the 
color-singlet model, (3.1) : the relativistic corrections of $O(v^{2})$ 
and the PQCD radiative corrections of $O(\alpha_{s})$ relative to the 
lowest order result shown in (3.1).
We briefly summarize both types of corrections in the rest of this
subsection, since they have to be included in principle 
for consistency, when one
includes the color-octet mechanism in many cases.

The relativistic corrections to the $\gamma-$gluon fusion was studied 
by Jung {\it et al.} \cite{jung}. They found that relativistic corrections
are important for high $z > 0.9$ at EMC energy ($\sqrt{s_{\gamma p}} \simeq
14.7$ GeV). Since it mainly affects the high $z$ region only, we neglect
the relativistic corrections, keeping in mind that it enhances the cross
section at large $z > 0.9$.  

The radiative corrections to the $J/\psi$ photoproduction is rather
important in practice.  
This calculation has been done recently in Ref.~\cite{kramer}, and 
the scale dependence of the lowest order result ($Q^{2}$ in the structure 
function in Eq. (3.10) ) becomes
considerably reduced. For EMC energy region, the $K$ factor is rather large, 
$K \sim 2$. For HERA, it depends on the cuts in
$z$ and $P_{T}^2$.  
We include the radiative corrections in terms of a $K$ factor suitable 
to the energy range we consider.
Another consequence of the radiative corrections to
the color-singlet $J/\psi$ photoproduction is that the PQCD 
becomes out of control for $z > 0.9$ at EMC energy. For HERA, one gets
reasonable results in PQCD when one imposes the following cuts in $z$ and 
$P_{T}^{2}$ : $z < 0.8$ and $P_{T}^{2} > 1~{\rm GeV}^2$ .
Thus, it is not too much of sense to talk about
the $z$ or $p_T$ distributions for such $z$ region in PQCD. 
One has to introduce cuts in $z$ as well as in $p_T$. Following 
the Ref.~\cite{kramer}, we adopt the following sets of cuts :
\begin{eqnarray}
z < 0.9, & ~~~~~& {\rm for ~EMC},
\\
0.2 < z < 0.8 & ~~~~~& {\rm for ~HERA}.
\end{eqnarray}
At HERA energies, the lower cut in $z (z > 0.2)$ is employed in 
order to reduce  backgrounds from the
resolved photon process and the $b$ decays into $J/\psi$. 
For these cuts, the $K$ factor is approximately $K \simeq 1.8$ both
at HERA and the fixed target experiments.  
We include these radiative corrections to the
subprocess (3.1) in Sec.~
\ref{subsec:photo} by setting $K \simeq 1.8$.

\vspace{.3in}

\subsection{Color-octet  contributions to $2 \rightarrow 1$ subprocesses 
}
\label{subsec:3b}

Let us consider color-octet contributions to the $2 \rightarrow 1$ 
subprocesses via
\begin{equation}
\gamma(k) + g^*_a(g) \rightarrow (c\bar{c})[^{2S+1}L^{(8b)}_J](P),
\end{equation}
followed by $(c\bar{c})_{8}$ fragmenting into $J/\psi$ with emission of
soft gluons.
This subprocess occurs at $O(\alpha \alpha_{s} v^{7})$.
Here, $a, b$ are color indices for the initial gluon and the final 
color-octet $c\bar{c}$ state, and we are interested in $S=L=J=0$ and 
$S=L=1, J=0,1,2$.  
There are 2 diagrams representing the vertex, as shown in Fig.~ 2.
Here we consider the process where only the gluon is  off-shell.
Following previous convention,
we first write the matrix ${\cal O}$ related to this effective vertices.
\begin{eqnarray}
{\cal O}(P,q)
&=&
\frac{ee_cg_s\delta^{ab}}{\sqrt{2}}
\left[
\not{\epsilon}^\gamma
\frac {\frac{\not{P}}{2}+\not{q}-\not{k}+M_c}
{(\frac{P}{2}+q-k)^2-M_c^2}
\not{\epsilon}^g
+
\not{\epsilon}^g
\frac {\frac{\not{P}}{2}+\not{q}-\not{g}+M_c}
{(\frac{P}{2}+q-g)^2-M_c^2}
\not{\epsilon}^\gamma
\right].
\end{eqnarray}
With this matrix ${\cal O}$
we can derive the effective vertices for the 
$\gamma g (c\bar{c})^{2S+1}L_J^{(8)}$
as
\begin{eqnarray}
{\cal M}^\prime(^1S_0^{(8)})&=& 4i\frac{ee_cg_s}{g^2-4M_c^2}
\delta^{ab} \epsilon^{\mu\nu\kappa\lambda} 
\epsilon^\gamma_\mu \epsilon^g_\nu P_\kappa k_\lambda,
\\
{\cal M}^\prime(^3S_1^{(8)})&=& 0,
\\
{\cal M}^\prime(^3P_0^{(8)})&=&
\frac{2e e_c g_s \delta^{ab}}{\sqrt{3}M_c}
\left(\frac{g^2-12M_c^2}{g^2-4M_c^2}\right)
\left(g^{\mu\nu}+2\frac{P^\mu k^\nu}{g^2-4M_c^2}\right)
\epsilon^\gamma_\mu
\epsilon^g_\nu,
\\
{\cal M}^\prime(^3P_1^{(8)})&=&
\frac{\sqrt{2}e e_c g_s\delta^{ab}}{M_c^2(g^2-4M_c^2)}
\nonumber\\
&\times &
\left(
      g^2\epsilon^{\mu\nu\alpha\tau}
     +{2k_\kappa}
       \frac{ g^2
               (  P^\mu\epsilon^{\nu\alpha\kappa\tau}
                 -P^\nu\epsilon^{\mu\alpha\kappa\tau}
                )
        +4g^\nu M_c^2\epsilon^{\mu\alpha\kappa\tau}
       } {g^2-4M_c^2} 
\right)
\epsilon_{\alpha}(J_z)
\epsilon^\gamma_\mu
\epsilon^g_\nu
P_\tau,
\\
{\cal M}^\prime(^3P_2^{(8)})&=&
\frac{16e e_c g_s\delta^{ab}}{(g^2-4M_c^2)}M_c
\left(
g^{\mu\alpha} g^{\nu\beta}
+2k^\alpha
\frac{
k^\beta g^{\mu\nu} + P^\mu g^{\nu\beta} - k^\nu g^{\mu\beta}
}
{g^2-4M_c^2}
\right)
\epsilon_{\alpha\beta}(J_z)
\epsilon^\gamma_\mu
\epsilon^g_\nu.
\end{eqnarray}
Since $J/\psi$ can be produced
via the $2\rightarrow 1$ subprocesses with this effective vertices
we can obtain the $2\rightarrow 1$ color octet contribution
by using the following average squared amplitudes as \footnote{Our results
agree with those obtained in Refs.~\cite{kramer2} \cite{fleming2}.
Note, however, that our convention of the invariant matrix is different from
theirs.}
\begin{eqnarray}
\overline{\sum}|{\cal M}^\prime(^{1}S_0^{(8)})|^2&=&2(ee_cg_s)^2,
\\
\overline{\sum}|{\cal M}^\prime(^{3}S_1^{(8)})|^2&=&0,
\\
\overline{\sum}|{\cal M}^\prime(^{3}P_0^{(8)})|^2&=&\frac{6}{M_c^2}
(ee_cg_s)^2,
\\
\overline{\sum}|{\cal M}^\prime(^{3}P_1^{(8)})|^2&=&0,
\\
\overline{\sum}|{\cal M}^\prime(^{3}P_2^{(8)})|^2&=&
\frac{8}{M_c^2}(ee_cg_s)^2.
\end{eqnarray}
The photoproduction cross section of the  $J/\psi$
production via $2\rightarrow 1$ process can be obtained from (\ref{eq:sigma}), 
assuming heavy quark spin symmetry
 :
\begin{eqnarray}
&&\sigma\left(\gamma+p\rightarrow (c\bar{c})^{(8)}\rightarrow \psi\right)
\nonumber\\
&&\hskip 2cm
=\frac{7\pi(ee_cg_s)^2}{64M_c^5}\left[xf_{g/p}(x)\right]_{x=4M_c^2/s}
\left(
      \frac{\langle 0|{\cal O}^{\psi}(^3P_0^{(8)})|0\rangle}{M_c^2}
     +\frac{\langle 0|{\cal O}^{\psi}(^1S_0^{(8)})|0\rangle}{7}
\right).
\end{eqnarray}
Since $\hat{\sigma} \propto \delta ( 1 -z)$, this $2\rightarrow 1$ 
color-octet subprocesses contribute to the elastic peak in the 
$J/\psi-$photoproductions.  It is timely to recall that the color-singlet
model with relativistic corrections still underestimate the cross section
for $z \ge 0.9$ by an appreciable amount \cite{jung}. 
As $z \rightarrow 1$, the final state gluon in the $\gamma-$gluon fusion 
becomes softer and softer, although this does not cause any infrared 
divergence in the transition matrix element.
Therefore, it would be more meaningful to factorize the effect of this 
final soft gluon 
into the color-octet matrix elements, $\langle O_{8}^{\psi} 
(^{1}S_{0}) \rangle$ and  $\langle O_{8}^{\psi} (^{3}P_{J} ) \rangle$. 
The color-octet $^{1}S_{0}$ and 
$^{3}P_{J}$ states might reduce the gap between the color-singlet prediction
and the experimental value of $d\sigma / dz$ for $0.9 \le z \le 1$.  

\vspace{.3in}

\subsection{Color-octet contributions to the $2\rightarrow 2$ subprocesses}
\label{subsec:3c}

The color-octet $2 \rightarrow 1$ subprocess (3.21) considered in the previous 
subsection not only contribute to the elastic peak of the $J/\psi$ 
photoproduction, but it also contributes  
to the resolved photon processes at $O(\alpha \alpha_{s}^{2} v^{7})$, 
shown in Fig.~3, where the initial partons
can be either gluon or light quarks ($q = u,d,s$). 
These diagrams are suppressed by $v^4$ but enhanced by $1/\alpha_s$, relative
to the resolved photon process in the color-singlet model. Also, the jet 
structures from Fig.~3  are different from that from the resolved photon 
process in the color-singlet model.  They can enhance the high-$p_T 
J/\psi$'s, which might be relevant to HERA energy.  
 This can be a background to the 
determination of gluon distribution function of a proton, if the cross 
section is appreciable.  The resolved photon process in the color-singlet 
model is dominant over the $\gamma-$gluon fusion in the lower $z$ region,
$z < 0.2$, and it can be discarded by a suitable cut on $z$. 
Since the color-octet contribution to the resolved photon process has not 
been studied in the literature, we address this issue in the following.
When one considers Fig.~3, one has to include Fig.~1 with $(c \bar{c})_8$
simultaneously,
since both are the same order of $O(\alpha \alpha_{s}^{2} v^{7})$.
This diagram is the same as the color-singlet case except for the color
factor of the $(c \bar{c})$ state.

It is straightforward, although lengthy, 
to calculate the amplitudes for the processes shown in Figs.~1 and 3.  
Using REDUCE in order to the spinor algebra in a symbolic manner, 
we can get the averaged ~${\cal M}$~ squared for various ~$2\to 2$~processes.
Since the full expressions are rather involved, they are shown in
Appendix A.

Another  color-octet $(c\bar{c})(^{3}S^{(8)}_{1})$ contribution to the
$J/\psi-$photoproductions comes from the Compton scattering type subprocesses
(see Fig.~4) :
\begin{equation}
\gamma (k,\epsilon) + q(p_{1}) \rightarrow (c\bar{c})(^{3}S^{(8a)}_{1})
(P,\epsilon^{*}) + q(p_{2}),
\end{equation}
where $P$ and $\epsilon^{*}$ are the four momentum and the polarization vector
of the  $^{3}S_{1}$ color-octet state, and $a$ is its color index.
This subprocess, if important, can be a background to the determination of
the gluon distribution function in a proton, since it is initiated by
light quarks.
>From the naive power counting, however, we infer this subprocess occurs at
$O(\alpha \alpha_{s}^{2})$ in the coupling constant expansion, and also
suppressed by $v^4$ compared to the color-singlet contribution (3.1)
due to its color-octet nature.  Thus, this subprocess is expected to be
negligible.

One can actually quantify this argument by explicitly evaluating the
Feynman diagrams shown in Fig.~4.
The effective vertex for $q \bar{q} \rightarrow
(c\bar{c})(^{3}S^{(8a)}_{1})$
is given by (Fig.~5) \cite{cho}
\begin{equation}
{\cal M}^\prime
(q(p_{1}) \bar{q} (p_{2}) \rightarrow (c\bar{c})(^{3}S^{(8a)}_{1})) =
{4 \pi \alpha_{s} \over 2 M_{c}}~\bar{v}(p_{2}) \gamma^{\mu} T^{a} u(p_{1})~
\epsilon_{\mu}^{*}(p_{1}+p_{2}, S_{z}),
\end{equation}
where $\epsilon_{\mu}^{*}$ is the polarization of the produced spin-1 color
octet object.
Using this effective vertex, one can calculate the amplitude for the
Feynman diagrams shown in Fig.~4 :
\begin{eqnarray}
{\cal M}^\prime
(\gamma q \rightarrow (c\bar{c})(^{3}S^{(8a)}_{1}) q) & = &
-{g_{s}^{2} e e_{q} \over 2 M_c}~
\bar{u}(p_{2}) ~
\left[ \not{\epsilon}^{*}(P,S_{z}) T_{a} ~{(k+p_{1}+M_{c}) \over
(k+p_{1})^{2} -M_{c}^2}~\not{\epsilon}_\gamma
\right.   \nonumber\\
&&\hskip 3cm + \left.   \not{\epsilon}_{\gamma} {(p_{1}-P + M_{c})
\over  (p_{1}-P)^{2} - M_{c}^2}
\not{\epsilon}^{*}(P,S_{z}) T_{a} \right]~u(p_{1}).
\end{eqnarray}
where $ee_q$ is the electric charge of the light quark
inside proton($q=u, d, s$).
The averaged ${\cal M}^2$ for the color-octet $^{3}S_1$ state is given by
\begin{equation}
\overline{\sum}| {\cal M}^\prime(\gamma q \rightarrow
(c\bar{c})(^{3}S^{(8)}_{1}) q) |^{2}
=  -\frac{2}{3M_c^2}(g_s^2 e e_q)^2
(\frac{\hat{s}}{\hat{u}}+ \frac{\hat{u}}{\hat{s}}+8\frac{M_c^2 
\hat{t}}{\hat{s} \hat{u}}).
\end{equation}
This completes our discussions on the color-octet $2 \rightarrow 2$ 
subprocess for $J/\psi$ photoproductions.

\vspace{.3in}

\section{Numerical results}
\label{sec:four}

\subsection{$J/\psi$ photoproductions at fixed targets and HERA}
\label{subsec:photo}

Now, we  are ready to show the numerical results using the analytic
expressions obtained in the previous section.  Let us first summarize the 
input parameters and the structure functions we will use in the 
following.  The results are quite sensitive to the numerical values of 
$\alpha_s$ and $m_{c}$ and the factorization scale $Q$. 
We shall use $\alpha_{s} (M_{c}^{2}) = 0.3$, $m_{c} = 1.48$ GeV and 
$Q^{2} = (2 m_{c})^2$. For the structure functions, we use 
the most recent ones, MRSA \cite{mrsa} and CTEQ3M \cite{cteq3}, 
which incorporate the new data
from HERA \cite{exp:hera}, on the lepton asymmetry in $W-$boson production 
\cite{exp:lepton} and on the difference in Drell-Yan cross sections
from proton and neutron targets \cite{exp:DY}.  
For the $2 \rightarrow 1$, we show results using both structure functions.
For the $2 \rightarrow 2$ case, we show the results with the CTEQ3M structure 
functions only, since the MRSA structure functions yield more or less the
same results within $\sim 10 \%$ or so.

Let us first consider the $J/\psi$ photoproduction via the color-octet 
$2 \rightarrow 1$ subprocess considered in Sec.~\ref{subsec:3b}.
Since the subprocess cross section (3.33) vanishes except at $z=1$, 
one can infer that it contributes to the $J/\psi$ photoproductions 
in the forward direction ($z \sim 1, P_{T}^{2} \simeq 0$). 
In Figs.~6 (a) and (b), we show the $J/\psi$ photoproduction cross section
in the forward direction as well as the data from the fixed target
experiments and the preliminary data from H1 at HERA, respectively.
In each case, the upper and the lower curves define the region allowed 
by the relation (1.4) for two color-octet matrix elements, 
$\langle 0 | O_{8}^{\psi} ({^{3}S_{1}}) | 0 \rangle$ and $\langle 0 
| O_{8}^{\psi} ({^{3}P_{0}}) | 0 \rangle$.   
In case of fixed target experiments, it is 
usually characterized by $z>0.9$, with the remainder being associated
with the inelastic $J/\psi$ photoproduction.  According to this criterion,
the experimental value of $\sigma_{\rm exp} (\gamma + p \rightarrow J/\psi
+ X)$ contains contributions from inelastic production of $J/\psi$'s.
Thus, the data should lie above the predictions from the color-octet 
$2 \rightarrow 1$ subprocess, (3.21).  Fig.~6 (a) shows that the situation
is opposite to this expectation.  Color-octet contributions are larger
than the data, which indicates that the numerical values of the color-octet
matrix elements are probably too large.   
At HERA, one has the elastic $J/\psi$ photoproduction data, which can be 
identified with  the color-octet $2 \rightarrow 1$ subprocess.
By saturating the relation 
relation (1.4) by either color-octet matrix element, we get the 
$J/\psi$ photoproduction cross section in the forward direction 
(Fig.~6 (b)).  We observe again that the color-octet contribution with (1.4)
overestimates the cross section by a large amount.  
This disagreement can arise from two sources : (i) the radiative 
corrections to $p \bar{p} \rightarrow J/\psi + X$, which were ignored in 
Ref.~\cite{cho2} is important, and/or (ii) the heavy quark spin symmetry for 
$\langle 0 | O_{8}^{\psi} ({^{3}P_{J}}) | 0 \rangle \approx 
(2J+1)~\langle 0 | O_{8}^{\psi} ({^{3}P_{0}}) | 0 \rangle$ may not be
a good approximation. Although the heavy quark spin symmetry relation 
is used quite often in heavy quarkonium physics, it may be violated 
by a considerable amount \cite{ko}.  

Recently, Fleming {\it et al.} performed the $\chi^2$ fit to the 
available fixed target experiments and the HERA data independently, and 
found that 
\begin{equation}
\langle 0 | O_{8}^{\psi} (^{1}S_{0}) | 0 \rangle + {7 \over M_{c}^2}~
\langle 0 | O_{8}^{\psi} (^{3}P_{0}) | 0 \rangle  = 
(0.020 \pm 0.001)~{\rm GeV}^3,
\end{equation}
using the MRSA$^{(')}$, and CTEQ3M structure functions with $\alpha_{s}
(2 M_{c}) = 0.26$ and $M_{c} = 1.5$ GeV.
This determination is not compatible with the relation (1.4), since the 
resulting $\langle O_{8}(^{3}P_{0}) \rangle$ is negative.
This is another way to say that the determination of the color-octet 
matrix elements from the $J/\psi$ productions at the Tevatron may not be 
that reliable. In fact, this is not very surprising, since the radiative
corrections to the lowest-order color-singlet contributions to 
$J/\psi$ hadronic productions are not included yet.  

Next, we consider $J/\psi$ photoproductions through $2 \rightarrow 2$
parton-level subprocesses.  
As discussed at the end of Sec.~\ref{subsec:3a}, the PQCD corrections
to the lowest order $\gamma + g \rightarrow J/\psi + g$ is not under
proper control for $z> 0.9$. Therefore, we impose a cut $z < 0.9$ at
EMC energy, $\sqrt{s_{\gamma p}} = 14.7$ GeV,  and 
at HERA with $\sqrt{s_{\gamma p}} = 100$ GeV, 
we impose cuts on $z$ and $P_{T}^2$ \cite{kramer} :
\[
0.2 < z < 0.8, ~~~~~~~~~~~~~~~~~~~P_{T}^{2} > 1~{\rm GeV}^2.
\]
In both cases, we set $K \simeq 1.8$. 

In Figs.~7 (a) and (b), we show the $d\sigma / dz$  distributions of 
$J/\psi$ at EMC (NMC) and HERA along with the corresponding data. 
In both cases, the color-octet ${^3S_1}$ contribution
(Compton scattering type) is negligible in most regions of $z$, and thus 
can be safely neglected. 
The thick dashed and the thin dashed  curves correspond to the cases where 
the relation (1.4) is saturated by $O_{8}(^{3}P_{J})$ and 
$O_{8}(^{1}S_{0})$, respectively.  The thick and the thin solid curves 
represent the sum of the color-singlet and the color-octet contributions, 
in case that the relation (1.4) is saturated by $\langle 0 | 
{\cal O}_{8}^{\psi}
({^1S_0}) | 0 \rangle$  and  $\langle 0 | {\cal O}_{8}^{\psi}({^3P_0}) | 0 
\rangle$, respectively. In either case, we observe that 
the color-octet ${^1S_0}$ and ${^3P_J}$ contributions begin to dominate the 
color-singlet contributions for $z > 0.6$, and become too large for 
high $z$ region considering we have not added the enhancements at high $z$ 
due to the relativistic corrections.
Thus, this behavior of rapid growing at high $z$ does not agree with the 
data points at EMC and HERA, if we adopt the determination (1.4) by Cho and
Leibovich  \cite{cho2}.

In Figs. ~8 (a) and (b), we show the $P_{T}^2$
distributions of $J/\psi$ at EMC and HERA, respectively.  
We find that the color-octet contributions through $2 \rightarrow 2$ 
subprocesses become dominant over the color-singlet contributions for most 
$P_{T}^2$ region.
Also, the color-octet contributions from ${^1S_0}$ and ${^3P_J}$ are  more 
important than the charm  quark  fragmentation considered by Godbole 
{\it et al.} \cite{godbole}.
However, this situation may be due to the overestimated color-octet matrix
elements as alluded in the previous paragraph.   

In Fig.~9, we show the inelastic $J/\psi$ photoproduction cross section as
a function of $\sqrt{s_{\gamma p}}$ with the cut, $ z < 0.8$ and 
$P_{T}^{2} > 1~{\rm GeV}^2$.  Again, the color-octet ${^3S_1}$ contribution is 
too small, and thus not shown in the Figure.  
Here again, the color-octet ${^1S_0}$ and ${^3P_J}$ contributions via 
$2 \rightarrow 2$ subprocesses  (Fig.~3) dominate the color-singlet 
contribution, if the relation (1.4) is imposed.  Although the total seems
to be in reasonable agreement with the preliminary H1 data, direct comparison
may be meaningful only if the cascade $J/\psi$'s from $b$ decays have been
subtracted out.  There are also considerable amount of uncertainties coming
from $M_c$ and $\alpha_s$.  Therefore, it is sufficient to say that the 
color-octet ${^1S_0}$ and ${^3P_J}$ state dominate the singlet contribution
to $J/\psi$ photoproduction, if the relation (1.4) is imposed. 

\subsection{Digression on $B \rightarrow J/\psi +X$}
\label{subsec:btocc}

Finding that the color-octet matrix elements determined from $J/\psi$
productions at the Tevatron seem to be  too large  in case of
$J/\psi$ photoproductions at fixed target experiments and HERA,
it is timely to consider the color-octet contributions to inclusive
$B$ meson decays into $J/\psi + X$ again.  In Ref.~\cite{ko},
a new factorization formula is derived for $B \rightarrow J/\psi + X$ :
\begin{eqnarray}
\Gamma (B \rightarrow J/\psi + X) & = & \left( { \langle 0 | O_{1}^{J/\psi}
(^{3}S_{1}) | 0 \rangle \over 3 M_{c}^2} - 
{ \langle 0 | P_{1}^{J/\psi} (^{3}S_{1}) | 0 \rangle \over 9 M_{c}^4}
\right) ~ (2C_{+} - C_{-}
)^{2} \left( 1 + {8 M_{c}^{2} \over M_{b}^2} \right)~\hat{\Gamma}_{0}
\nonumber        \\
& + & {3 \langle 0 | O_{8}^{J/\psi}(^{1}S_{0}) | 0 \rangle
\over 2 M_{c}^2}~(C_{+} + C_{-})^{2}~\hat{\Gamma}_{0} 
\\
& + & { \langle 0 | O_{8}^{J/\psi}(^{3}P_{1}) | 0 \rangle
\over  M_{c}^4}~(C_{+} + C_{-})^{2}~\left( 1 + {8 M_{c}^{2} \over M_{b}^2}
\right)~\hat{\Gamma}_{0},
\nonumber
\end{eqnarray}
with
\begin{equation}
\hat{\Gamma}_{0} \equiv |V_{cb}|^{2} \left( {G_{F}^{2} \over 144 \pi}
\right) M_{b}^{3} M_{c} \left( 1 - {4 M_{c}^{2} \over M_{b}^2} \right)^{2}.
\end{equation}
Two numbers $C_{+} (M_{b}) \approx 0.87$ and $C_{-} (M_{b}) \approx 1.34$ 
are the Wilson coefficients of the $| \Delta B | = 1$ effective weak 
Hamiltonian.  
Using the relation (1.4), we estimate the above branching ratio
to be (for $\alpha_{s}(M_{\psi}^2) = 0.28$ in Ref.~\cite{ko}) 
\begin{equation}
(0.42 \% \times 12.8 )~< B (B \rightarrow J/\psi + X)  < 
~ (0.42 \% \times 13.8) 
\end{equation}
which is larger than the recent CLEO data by an order of magnitude 
\footnote{Even if we use the new determination (4.1) by Fleming {\it et al.} 
\cite{fleming2}, we still get a large branching ratio :
\begin{equation}
(0.42 \% \times 3.45 )~< B (B \rightarrow J/\psi + X)  <
~ (0.42 \% \times 5.45),
\end{equation}
although the discrepancy gets milder than the case (4.4).}: 
\begin{equation}
B_{\rm exp} (B \rightarrow J/\psi + X)  = (0.80 \pm 0.08) \%.
\end{equation}
The situation is the same for $B \rightarrow \psi^{'} + X$.
This is problematic, unless this large color-octet contributions are canceled
by the color-singlet contributions of higher order in $O(\alpha_{s})$ which 
were not included in Ref.~\cite{ko}.   
If there are no such fortuitous cancelations among various color-octet and 
the color-singlet contributions,
this disaster could be attributed to the relation (1.4) being   
too large compared to the naive velocity scaling rule in NRQCD,
as noticed in  Ref.~\cite{cho2}.
It seems to be crucial to include the higher order 
corrections of $O(\alpha_{s}^{4})$ for the color-singlet  $J/\psi$ 
productions at the Tevatron, which is still lacking in the literature.

\vspace{.3in}

\section{Conclusion}
\label{sec:five}

In summary, we considered the color-octet contributions to (i) the subprocess 
$\gamma  + p \rightarrow J/\psi + X$ through $\gamma g \rightarrow 
(c\bar{c})_{8} ({^1S_0}~{\rm and}~{^3P_J})$  and the subsequent evolution of 
$(c\bar{c})_8$  into a physical $J/\psi$  with $z \approx 1$ and $P_{T}^2 
\approx 0$, (ii) the subprocesses $\gamma + g ({\rm or}~q) \rightarrow 
(c\bar{c})_{8} ({^1S_0}~{\rm or}~{^3P_J}) + g ({\rm or}~ q)$.  
These are compared with (i) the measured $J/\psi$ photoproduction cross 
section in the forward direction, and (ii) the $z$ distributions of $J/\psi$ 
at  EMC and HERA, and the preliminary result on the inelastic $J/\psi$ 
photoproduction total cross section at HERA. One finds that the relation (1.4) 
color-octet   lead to too large contributions of the color-octet ${^1S_0}$ and 
${^3P_J}$ states to the above observables.  Especially, the first two 
observables contradict the observation.   This is also against the naive
expectation that the color-octet contribution may not be prominent as in the
case of the $J/\psi$ hadroproductions, since they are suppressed by $v^4$
(although enhanced by one power of $\alpha_s$) relative to the color-singlet
contribution.  
It is also pointed out that the same is true of the process $B \rightarrow
J/\psi +X$, in which the relation (1.4) predicts its branching ratio to be
too large by an order of magnitude compared with the data.

Therefore, one may conclude  that the color-octet matrix
elements involving $c\bar{c}_{8}({^{1}S_{0}}, {^{3}P_{J}})$ might be 
overestimated by an order of magnitude.  Since the relation (1.4) has been 
extracted by fitting the $J/\psi$ production at the Tevatron to the lowest
order color-singlet and the color-octet contributions,
it may be changed when one considers the radiative corrections to the 
lowest order color-singlet contributions.

\vspace{.2in}

While we were finishing our work, there appeared two papers considering
the same subject \cite{kramer2} \cite{fleming2}.  
Our results agree with these two works where they overlap.

\vspace{.3in}

\acknowledgements
This work was supported in part by KOSEF through CTP at Seoul National 
University. P.K. is supported in part by the Basic Science Research 
Program, Ministry of Education, 1995, Project No. BSRI--95--2425, and 
in part  by   NON DIRECTED RESEARCH FUND, Korea Research  Foundations
(1995).

\appendix
\section{The invariant amplitude squared for the $2 \rightarrow 2$ 
subprocesses considered in Sec.~III.C.}

In this Appendix, we give explicit expressions for the invariant amplitude 
squared for the color-octet 
$2 \rightarrow 2$ subprocesses shown in Figs.~1 and 3.
The results were obtained using the symbolic manipulations with the aid of
REDUCE package.  
We checked that the following formulae in two independent ways :
the helicity amplitude and the covariant density matrix methods in order to
do the summations over gluon polarizations. Now both methods yield the same
results listed below
\footnote{
The original expressions (A1)-(A4) are modified, because of 
maltreatment of gluon polarizations of Fig.~3 with three gluon 
vertices.
The numerical results based on the correct expressions shown here were
already given in the Ref.~\cite{bernd}, and so we do not reproduce them here.

JL would like to thank Avto Tkabladze for pointing out this error
and checking that the above formulae are correct. JL also would like to thank
B. Kniehl and G. Kramer for using the corrected formulae for their analysis
in Ref.~\cite{bernd}
where they reproduced exactly the same numerical results as
in Ref~\cite{kramer2}.
} : 
\begin{eqnarray}
\left(
\frac{d\sigma}{d\hat{t}}
\right)
(\gamma+g\to (c\overline{c}){^{2S+1}L_J^{(8)}}+g \to J/\psi+X)
=
\frac{(e_0 e_c g_s^2)^2}{16\pi\hat{s}^2}
\frac{
\langle O^{J/\psi}(^{2S+1}L_J^{(8)})\rangle
}{(2J+1)M_c}
\times
f({^{2S+1}L_J^{(8)}})
\nonumber
\end{eqnarray}
\begin{eqnarray}
f(^1S_0^{(8)})&=&
\frac{
3\hat{s}\hat{u}
}
{
2\hat{t}
(\hat{s}+\hat{t})^2(\hat{t}+\hat{u})^2(\hat{u}+\hat{s})^2
}
\bigg[\;
\hat{s}^4+\hat{t}^4+\hat{u}^4+(2M_c)^8
\bigg]
\\
f(^3P_0^{(8)})&=&
\hat{s}\hat{u}
\bigg[
\;
\hat{s}^2(\hat{t}+\hat{u})^2
\left\{
2\hat{t}\hat{u}(\hat{t}+\hat{u})-3\hat{s}^2(2M_c)^2+\hat{t}\hat{s}\hat{u}
\right\}^2
+9(2M_c)^8
(\hat{t}+\hat{s})^2(\hat{s}+\hat{u})^2(\hat{u}+\hat{t})^2
\nonumber\\
&&
\hspace{1ex}
+\hat{u}^2(\hat{t}+\hat{s})^2
\left\{
2\hat{t}\hat{s}(\hat{t}+\hat{s})-3\hat{u}^2(2M_c)^2+\hat{t}\hat{s}\hat{u}
\right\}^2
\nonumber\\
&&
\hspace{1ex}
+\hat{t}^4(\hat{s}+\hat{u})^2
\left\{
5(\hat{t}+\hat{s})(\hat{t}+\hat{u})
+2\hat{t}(\hat{s}+\hat{u})+2(\hat{s}^2+\hat{u}^2)
\right\}^2
\bigg]
\nonumber\\
&&/
\hspace{1ex}
\bigg[
{
2M_c^2\hat{t}(\hat{s}+\hat{t})^4(\hat{t}+\hat{u})^4
(\hat{u}+\hat{s})^4
}
\bigg]
\\
f(^3P_1^{(8)})&=&
{3}
\bigg[
\;
(2M_c)^2 \hat{t}^2
\left\{ 
 \hat{s}^4(\hat{t}+\hat{u})^2\left(\hat{u}+(2M_c)^2\right)^2
+\hat{u}^4(\hat{t}+\hat{s})^2\left(\hat{s}+(2M_c)^2\right)^2
\right\}
\nonumber\\
&&
+(2M_c)^2 \hat{s}^2\hat{u}^2
\left\{
(\hat{t}+\hat{u})^2
\left(
(\hat{t}+\hat{s})^2+(\hat{t}-\hat{s})\hat{u}
\right)^2
+(\hat{t}+\hat{s})^2
\left(
(\hat{t}+\hat{u})^2+(\hat{t}-\hat{u})\hat{s}
\right)^2 
\right\}
\nonumber\\
&&
+(2M_c)^2 (\hat{s}^2-\hat{u}^2)\hat{t}^2
\left\{
(\hat{t}+\hat{u})(\hat{t}+\hat{s})-2\hat{s}\hat{u}
\right\}^2
\nonumber\\
&&
+2\hat{t}\hat{s}\hat{u}
\left((2M_c)^4-\hat{s}\hat{u}\right)^2
(\hat{s}^2(\hat{t}+\hat{u})^2+ \hat{u}^2(\hat{t}+\hat{s})^2
+ \hat{t}^2(\hat{s}+\hat{u})^2)
\bigg]
\nonumber\\
&&
/\bigg[
{
2M_c^2(\hat{s}+\hat{t})^4(\hat{t}+\hat{u})^4(\hat{u}+\hat{s})^4
}
\bigg]
\\
f({^3P_2^{(8)}})&=&f({^3P_A^{(8)}})
                 - \left(f({^3P_0^{(8)}})+f({^3P_1^{(8)}})\right)
\\
f(^3P_A^{(8)})&=&
\frac{3}
{
M_c^2\hat{t}(\hat{s}+\hat{t})^3(\hat{t}+\hat{u})^3(\hat{u}+\hat{s})^3
}
\bigg[
\;
  \hat{t}^6(\hat{s} + \hat{u})
\left\{
2(\hat{s}^2+\hat{u}^2) + 11\hat{s}\hat{u} 
\right\}
\nonumber\\
&&
+ \hat{t}^5
\left\{
4(\hat{s}^4+\hat{u}^4) 
+ 47\hat{s}\hat{u}(\hat{s}^2+\hat{u}^2) 
+ 70 \hat{s}^2\hat{u}^2 
\right\}
\nonumber\\
&&
+ \hat{t}^4(\hat{s} + \hat{u})
\left\{
   2(\hat{s}^4+\hat{u}^4) 
+ 61\hat{s}\hat{u} (\hat{s}^2+\hat{u}^2)
+ 75\hat{s}^2\hat{u}^2 
\right\}
\nonumber\\
&&
+ \hat{t}^3\hat{s}\hat{u}
\left\{
   47(\hat{s}^4+\hat{u}^4) 
+ 132\hat{s}\hat{u} (\hat{s}^2+\hat{u}^2)
+ 190 \hat{s}^2\hat{u}^2 
\right\}
\nonumber\\
&&
+ \hat{t}^2\hat{s}\hat{u} 
(\hat{s} + \hat{u})
\left\{
  25(\hat{s}^4+\hat{u}^4) 
+ 63\hat{s}\hat{u} (\hat{s}^2+\hat{u}^2)
+ 93\hat{s}^2\hat{u}^2 
\right\}
\nonumber\\
&&
+ \hat{t}\hat{s}\hat{u}
\left\{
   7(\hat{s}^6+\hat{u}^6) 
+ 38\hat{s}  \hat{u}  (\hat{s}^4+\hat{u}^4)
+ 78\hat{s}^2\hat{u}^2(\hat{s}^2+\hat{u}^2)
+ 98\hat{s}^3\hat{u}^3 
\right\}
\nonumber\\
&&
+ 7\hat{s}^2\hat{u}^2(\hat{s}+\hat{u}) 
(\hat{s}^2 + \hat{s}\hat{u} + \hat{u}^2)^2
\bigg]
\nonumber\\
\overline{\sum}
|{\cal M}^\prime|^2
&&(\gamma+q\rightarrow ~(c\bar{c})~(^1S^{(8)}_0)+q)=
-\frac{16}{3}(e e_c g_s^2)^2
\frac{\hat{s}^2+\hat{u}^2}{\hat{t}(\hat{s}+\hat{u})^2},
\\
\overline{\sum}
|{\cal M}^\prime|^2
&&(\gamma+q\rightarrow ~(c\bar{c})~(^3P^{(8)}_0)+q)=
-\frac{16}{9}(e e_c g_s^2)^2
\frac{(\hat{s}^2+\hat{u}^2)(\hat{t}-12M_c^2)^2}
     {\hat{t}(\hat{s}+\hat{u})^4M_c^2},
\\
\overline{\sum}
|{\cal M}^\prime|^2
&&(\gamma+q\rightarrow ~(c\bar{c})~(^3P^{(8)}_1)+q)=
-\frac{32}{3}(e e_c g_s^2)^2
\frac{(\hat{s}^2+\hat{u}^2)\hat{t}+16M_c^2 \hat{s}\hat{u}}
     {(\hat{s}+\hat{u})^4M_c^2},
\\
\overline{\sum}
|{\cal M}^\prime|^2
&&(\gamma+q\rightarrow ~(c\bar{c})~(^3P^{(8)}_2)+q)=
\nonumber\\
&&
\hskip -1cm
-\frac{32}{9}(e e_c g_s^2)^2
\frac{
(\hat{s}+\hat{u})^2 (\hat{t}^2+96M_c^4)
-2\hat{s}\hat{u}((\hat{s}+\hat{u}+4M_c^2)^2+8M_c^2(\hat{s}+\hat{u}))
}
{\hat{t}(\hat{s}+\hat{u})^4M_c^2}.
\end{eqnarray}
Here, the $e_c = 2/3$ and we have summed over the electric charges 
of light quarks ($q = u,d,s$) in the above expressions,  assuming 
$m_q = 0$.   

\newpage
\begin{center}
FIGURE CAPTIONS
\end{center}
\noindent
Fig.1
\hskip .3cm
{Feynman diagrams for the color-singlet and the 
color-octet subprocess
~$\gamma+ g \rightarrow (c\bar{c})_{1,8}({^{3}S_{1}}) + g$.
}
\\
\\
Fig.2
\hskip .3cm
{Feynman diagrams for the color-octet subprocess
~$\gamma+ g \rightarrow (c\bar{c})_{8}(^{1}S_{0}~{\rm or} 
~^{3}P_{J})$.
}
\\
\\
Fig.3
\hskip .3cm
{Feynman diagrams for the color-octet contribution
to the resolved photon 
$\gamma + g ({\rm or}~q) \rightarrow (c \bar{c})_{8}
(^{1}S_{0}~{\rm or} ~^{3}P_{J}) + g ({\rm or}~q)$.}
\\
\\
Fig.4
\hskip .3cm
{
Feynman diagrams for the color-octet 
$2\rightarrow 2$ subprocess, $\gamma + q \rightarrow (c\bar{c})_{8} ({^3S_1})
+ q$ with $q = u,d,s$.
}
\\
\\
Fig.5
\hskip .3cm
{A Feynman diagram for $q \bar{q} \rightarrow (c \bar{c})_{8}
(^3S_1)$.}
\\
\\
Fig.6(a)
\hskip .3cm
{
The cross sections for $\gamma + p \rightarrow J/\psi + X$
in the forward direction at
the fixed target experiments
as a function of $E_{\gamma}$.
The solid and the dashed curves were obtained using the CTEQ3M and the MRSA
structure functions.
Here, TOT$_s$ is the $^1S^{(8)}_0$ saturated curve and
TOT$_p$ is the $^3P^{(8)}_J$ saturated one. 
}
\\
\\
Fig.6(b)
\hskip .3cm
{
The cross sections for $\gamma + p \rightarrow J/\psi + X$
in the forward direction at
HERA 
as a function of the square root of $s_{\gamma p}$. 
The solid and the dashed curves were obtained using the CTEQ3M and the MRSA
structure functions.
Here, TOT$_s$ is the $^1S^{(8)}_0$ saturated curve and
TOT$_p$ is the $^3P^{(8)}_J$ saturated one. 
}
\\
\\
Fig.7(a)
\hskip .3cm
{
The differential cross sections $d\sigma / dz$ for
$\gamma + p \rightarrow J/\psi + X$  at EMC
as a function of $z\equiv E_{J/\psi}/ E_\gamma$.
The singlet contributions are in the thick dotted curve,
the color-octet $^1S_0$ contributions in the thick dashed curve
(with $\langle O_{8}^{\psi} (^1S_0) \rangle = 6.6 \times 10^{-2}
~{\rm GeV}^{3}$),
 and the color-octet $^3P_J$ contributions in the thin dashed curve
(with $\langle O_{8}^{\psi} (^3P_J) \rangle 
/M_c^2= 2.2 \times 10^{-2}
~{\rm GeV}^{3}$).
 The total is shown in the solid curve.
The relation (1.4) allows the region between two solid curves.
Here, TOT$_s$ is the $^1S^{(8)}_0$ saturated curve and
TOT$_p$ is the $^3P^{(8)}_J$ saturated one. 
}
\\
\\
Fig.7(b)
\hskip .3cm
{
The differential cross sections $d\sigma / dz$ for
$\gamma + p \rightarrow J/\psi + X$  at HERA
as a function of $z\equiv E_{J/\psi}/ E_\gamma$.
The singlet contributions are in the thick dotted curve,
the color-octet $^1S_0$ contributions in the thick dashed curve
(with $\langle O_{8}^{\psi} (^1S_0) \rangle = 6.6 \times 10^{-2}
~{\rm GeV}^{3}$),
 and the color-octet $^3P_J$ contributions in the thin dashed curve
(with $\langle O_{8}^{\psi} (^3P_J) \rangle
/M_c^2= 2.2 \times 10^{-2}
~{\rm GeV}^{3}$).
 The total is shown in the solid curve.
The relation (1.4) allows the region between two solid curves.
Here, TOT$_s$ is the $^1S^{(8)}_0$ saturated curve and
TOT$_p$ is the $^3P^{(8)}_J$ saturated one. 
}
\\
\\
Fig.8(a)
\hskip .3cm
{
The differential cross sections $d\sigma / dP_T^2$ for
$\gamma + p \rightarrow J/\psi + X$  at HERA 
as a function of $P_T^2$.
The singlet contributions in the thick dotted curve,
the color-octet $^1S_0$ contributions in the thick dashed curve
(with $\langle O_{8}^{\psi} (^1S_0) \rangle = 6.6 \times 10^{-2}
~{\rm GeV}^{3}$),
 and the color-octet $^3P_J$ contributions in the thin dashed curve
(with $\langle O_{8}^{\psi} (^3P_J) \rangle 
/M_c^2= 2.2 \times 10^{-2}
~{\rm GeV}^{3}$).
 The total is shown in the solid curve.
The relation (1.4) allows the region between two solid curves.
Here, TOT$_s$ is the $^1S^{(8)}_0$ saturated curve and
TOT$_p$ is the $^3P^{(8)}_J$ saturated one. 
}
\\
\\
Fig.8(b)
\hskip .3cm
{
The differential cross sections $d\sigma / dP_T^2$ for
$\gamma + p \rightarrow J/\psi + X$  at HERA
as a function of $P_T^2$ of $J/\psi$.
The singlet contributions in the thick dotted curve,
the color-octet $^1S_0$ contributions in the thick dashed curve
(with $\langle O_{8}^{\psi} (^1S_0) \rangle = 6.6 \times 10^{-2}
~{\rm GeV}^{3}$),
 and the color-octet $^3P_J$ contributions in the thin dashed curve
(with $\langle O_{8}^{\psi} (^3P_J) \rangle
/M_c^2= 2.2 \times 10^{-2}
~{\rm GeV}^{3}$).
 The total is shown in the solid curve.
The relation (1.4) allows the region between two solid curves.
Here, TOT$_s$ is the $^1S^{(8)}_0$ saturated curve and
TOT$_p$ is the $^3P^{(8)}_J$ saturated one. 
}
\\
\\
Fig.9
\hskip .3cm
{ Total inelastic $J/\psi$ photoproduction cross section for $z < 0.8$
as a function of the square root of $s_{\gamma p}$.
The singlet contributions in the thick dotted curve,
the color-octet $^1S_0$ contributions in the thick dashed curve
(with $\langle O_{8}^{\psi} (^1S_0) \rangle = 6.6 \times 10^{-2}
~{\rm GeV}^{3}$),
 and the color-octet $^3P_J$ contributions in the thin dashed curve
(with $\langle O_{8}^{\psi} (^3P_J) \rangle
/M_c^2= 2.2 \times 10^{-2}
~{\rm GeV}^{3}$).
 The total is shown in the solid curve.
The relation (1.4) allows the region between two solid curves.
Here, TOT$_s$ is the $^1S^{(8)}_0$ saturated curve and
TOT$_p$ is the $^3P^{(8)}_J$ saturated one. 
}
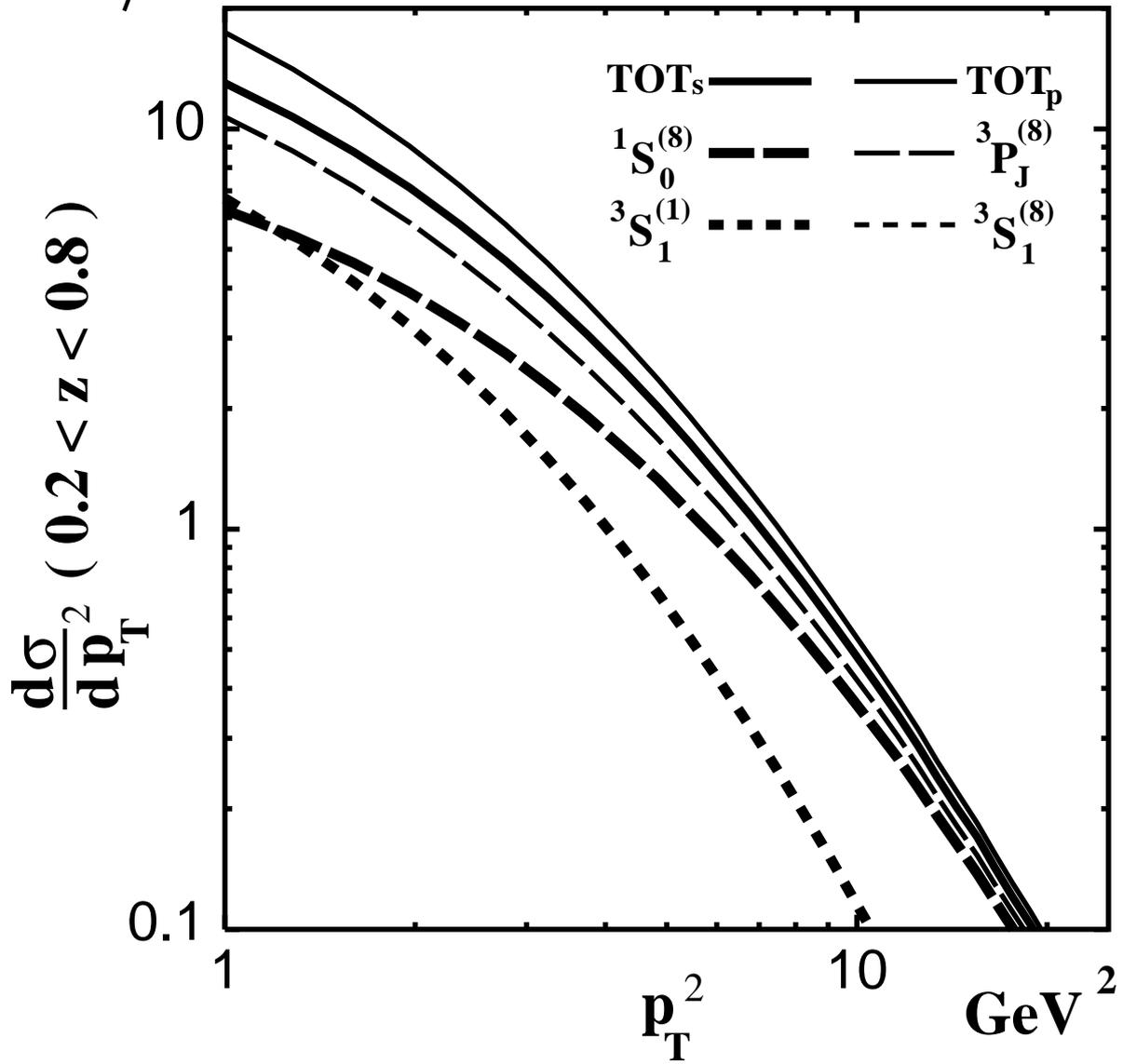
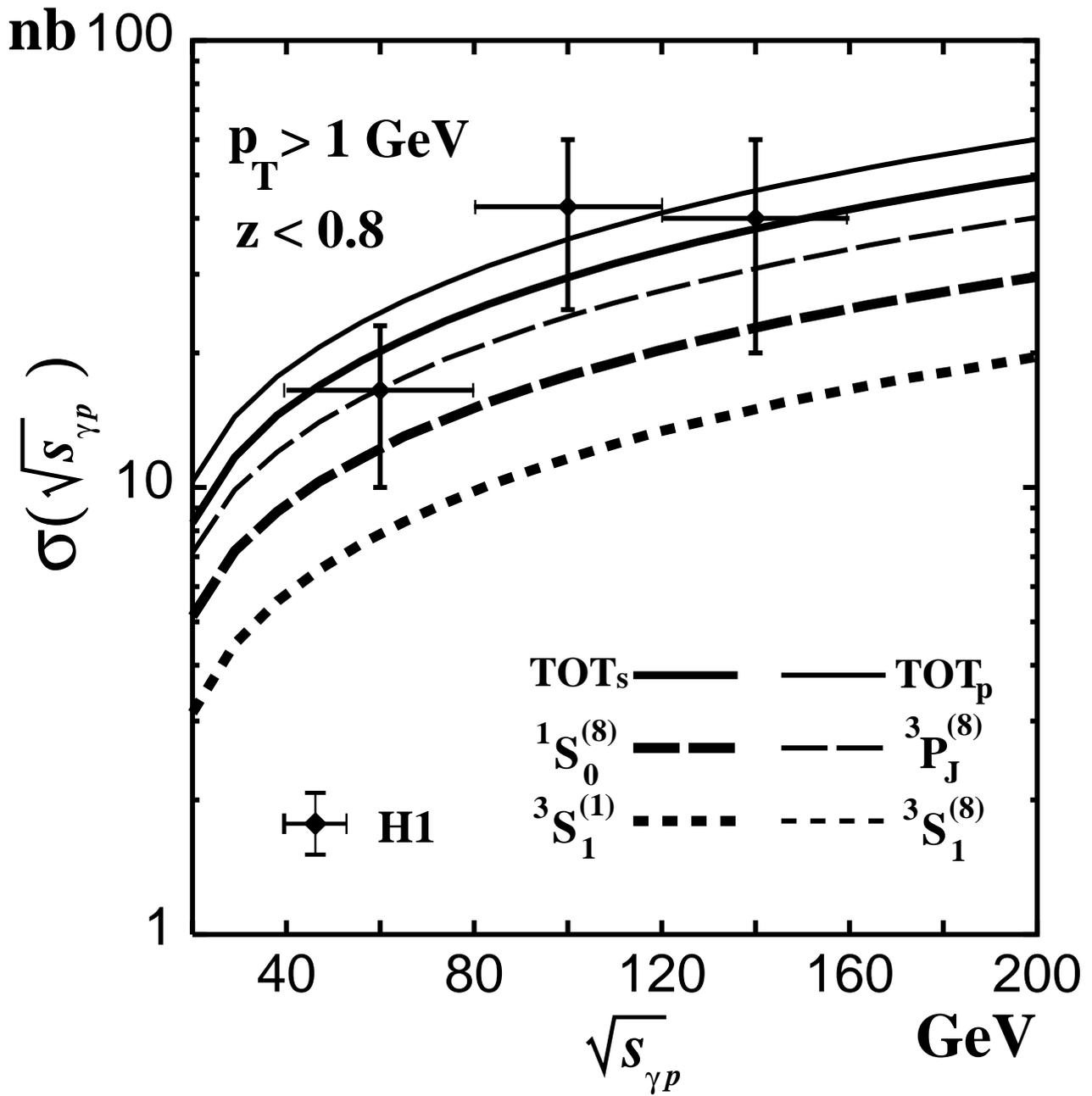
\end{document}